\documentclass[a4paper,fleqn,usenatbib]{mnras}

\usepackage{graphicx}
\usepackage{hyperref}
\usepackage{amssymb,amsmath}
\usepackage{breqn} 
\usepackage{bm}
\usepackage{dcolumn}

\usepackage{color}
\definecolor{purple}{rgb}{0.8,0.4,0.8}

\date{\today}
\pubyear{2018}

\title{Neutron star -- axion star collisions in the light of multi-messenger astronomy}

\author[T.~Dietrich,  F.~Day, K.~Clough, M.~Coughlin, J.~Niemeyer]
{Tim Dietrich$^1$,
Francesca Day$^2$,
Katy Clough$^3$,
Michael Coughlin$^4$,
Jens Niemeyer$^3$\\
${}^1$ Nikhef, Science Park, 1098 XG Amsterdam, The Netherlands\\
${}^2$ DAMTP, Centre for Mathematical 
Sciences, Wilberforce Road, Cambridge, CB3 0WA,
United Kingdom\\
${}^3$ Institut f{\"u}r Astrophysik, Georg-August 
  Universit{\"a}t, Friedrich-Hund-Platz 1, D-37077 G{\"o}ttingen, Germany \\
${}^4$ Division of Physics, Math, and Astronomy, California Institute of Technology, Pasadena, CA 91125, USA}

\begin{document}
\label{firstpage}
\pagerange{\pageref{firstpage}--\pageref{lastpage}}
\maketitle

\begin{abstract}
Axions are increasingly favoured as a candidate particle for the dark matter in galaxies, since they 
satisfy the observational requirements for cold dark matter and are theoretically well motivated.
Fluctuations in the axion field give rise to stable localised overdensities known as 
axion stars, which, for the most massive, compact cases, are potential neutron star mimickers. 
In principle, there are
no fundamental arguments against the  
multi-messenger observations of GW170817/GRB170817A/AT2017gfo arising from the merger 
of a neutron star with a neutron star mimicker, rather than from a binary neutron star.
To constrain this possibility and better understand the astrophysical signatures of a 
neutron star--axion star (NSAS) merger, 
we present in this work a detailed example case of a NSAS merger based on full 3D numerical relativity simulations,
and give an overview of the many potential 
observables - ranging from gravitational waves, 
to optical and near-infrared electromagnetic signals, 
radio flares, fast radio bursts, gamma ray bursts, 
and neutrino emission. 
We discuss the individual channels and estimate to which 
distances current and future observatories might be able 
to detect such a NSAS merger. Such signals 
could constrain the unknown axion mass and its couplings to standard baryonic matter,
thus enhancing our understanding of the dark matter sector 
of the Universe. 
\end{abstract}

\section{\label{sec:level1}Introduction}

The breakthrough discovery of 
GW170817~\cite{TheLIGOScientific:2017qsa} 
with the combined 
detection of the gamma-ray burst GRB170817A~\citep{Monitor:2017mdv} 
and the transient AT2017gfo~\citep{GBM:2017lvd} 
was the first coincident observation of gravitational 
waves (GWs) and electromagnetic (EM) waves from the same astrophysical 
source, and heralded a new era of multi-messenger astronomy. 
While there is good evidence that GW170817, GRB170817A, and
AT2017gfo were created by the coalescence and merger 
of two neutron stars (NSs)~\citep{Abbott:2018wiz,Abbott:2018exr}, 
it cannot yet be ruled out that
the observed GW and EM signals came from the merger of a NS with a NS-mimicker. 
As shown in~\cite{Cardoso:2017cfl,Sennett:2017etc} it is difficult 
to clearly distinguish NSs from exotic compact objects, 
e.g.~boson stars (BSs), with second generation GW detectors.

BSs are stable solitonic solutions to the coupled 
Einstein-Klein-Gordon equations, which describe a massive 
scalar field in the presence of gravity. Axion stars (ASs) are a particular kind of BS - real scalar 
field oscillotons with additional self interactions given by their non trivial field potential $V(\phi)$. 
Axions, although still unobserved, are theoretically well motivated: 
they explain the observed matter-anti-matter asymmetry via CP 
violation~\citep{Peccei:1977hh}, arise naturally
in string theory compactifications (see e.g.~\cite{Arvanitaki:2009fg}),
and are promising candidates for 
dark matter (DM) (see ~\cite{Marsh:2015xka} for a comprehensive review). ASs can form dynamically during the collapse of axion miniclusters in the early universe \citep{Hogan:1988mp,Kolb:1993zz} in a process similar to galactic core formation in ultra-light axion cosmologies \citep{Schive:2014dra,Veltmaat:2018dfz}, by wave condensation \citep{Levkov:2018kau}, or from non-standard primordial perturbations with enhanced small-scale power \citep{Widdicombe:2018oeo}. Whilst these scenarios generally predict the substantial majority of axionic dark matter to remain unbound, or bound in ASs in the low mass range, the high-mass tail of the AS mass fraction at low redshifts is widely unconstrained, motivating an exploration of observable signatures.

In this work we build on the results of~\cite{Clough:inprep} 
in which a large number of different NSAS configurations were 
simulated. Although restricted to head-on collisions the simulations 
showed that for NSAS systems close to the threshold 
of BH formation a large fraction of the bosonic and baryonic 
material can be ejected from the system and that a significant 
release of GW energy occurs during the collision and 
the post-merger phase. 
Here we present a detailed case study of the observables from a particular NSAS merger based on the full 3D numerical relativity simulations.
Using our simulation results, we also consider semi-analytically the potential conversion of axions 
to photons due to couplings to standard matter. 
Combined, these multi-messenger signals would lead to a unique signature for NSAS mergers. 

Note that to facilitate easier comparison with existing literature, 
we have used different unit systems for different 
multi-messenger channels, i.e., we employ geometric units for describing
the GW signal, $\rm cgs$-units for the kilonova and radio observations, 
and Planck units for the discussion of observables caused by the 
conversion of axions to photons.

\begin{figure}
\includegraphics[width=\columnwidth]{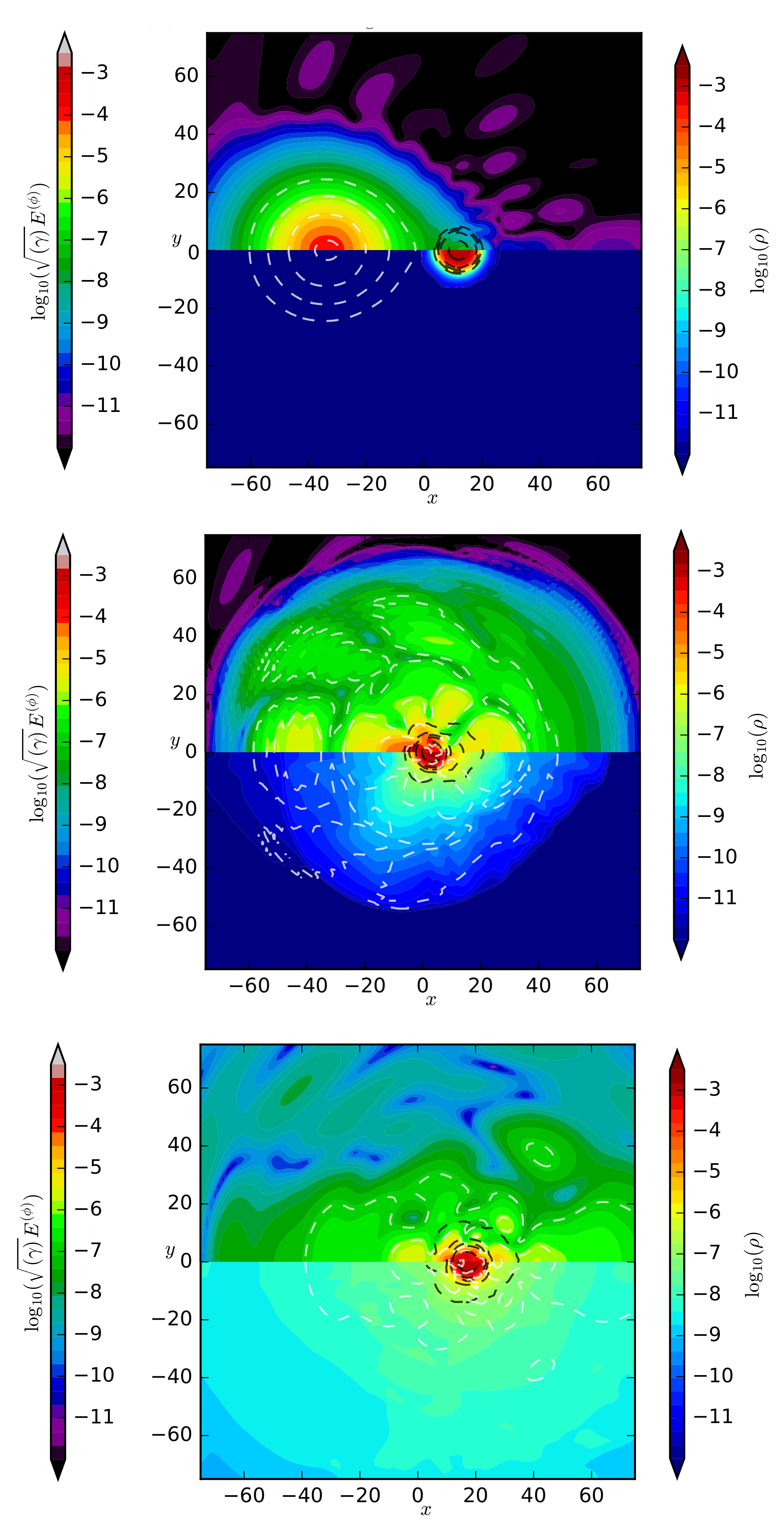}
\caption{Energy density of the axionic matter (top part of each panel) and the baryonic density (bottom part of each panel) for the times $t=9.3,10.8,17.6 \ {\rm ms}$.
We also include contour densities lines corresponding to 
$10^{-7},10^{-6},10^{5},10^{-4},10^{-3},10^{-2}$ for the 
bosonic energy density (white dashed lines) and baryonic energy density (black dashed lines).}
\label{fig:dynamics}
\end{figure}

\section{Configuration and merger dynamics}
\label{sec:dynamics}

Within this article we consider the head-on collision of 
the NSAS system NSAS$_{\rm 007}$ of~\cite{Clough:inprep}. 
For completeness, we review the configuration's key 
properties here (see~\cite{Clough:inprep} for a detailed discussion 
of the numerical methods and tests). 
The initial configuration is based on the superposition 
of a single isolated NS and AS separated by $260{\rm km}$. 
We solve the constraint equations to obtain a solution consistent 
with General Relativity~\citep{Dietrich:2018bvi,Clough:inprep}. 

The NS, with a gravitational mass of $\sim 1.38M_\odot$ 
in isolation, employs a piecewise-polytropic fit 
of the SLy Equation of State (EOS)~\citep{Douchin:2001sv,Read:2008iy}, 
which is in agreement with current
constraints inferred from~\cite{TheLIGOScientific:2017qsa}.
The AS employs the periodic cosine self interaction potential
$V(\phi) =  f_a^2 \mu^2 (1 - \cos({\phi}/{f_a}))$
with the axion decay constant $f_a$ (here set 
to $f_a=0.5  M_{pl}$, for negligible self-interactions) and the axion length scale
$\mu = m_a c / \hbar$ (here set to $GM_\odot/c^2$, of order of the NS radius). 
The physical axion mass is denoted by $m_a$ and for this system is of
the order of $10^{-10}\ {\rm eV}$.
We set the initial amplitude of the scalar field to 
$\phi_c\approx 0.014$, which results in an AS 
mass of about $0.36M_\odot$\footnote{\cite{Widdicombe:2018oeo} find that for 
$f_a=0.5 M_{pl}$ the formation of ASs in this mass range 
is one of the most favoured scenarios for their approximate model.}.

In Fig.~\ref{fig:dynamics} we show the axion energy density (top part of each panel)
and baryonic density (bottom part of each panel) during the simulation. 
At $t=9.3\ {\rm ms}$ (top panel), the AS is highly deformed 
due to the deep gravitational well of the NS, and bosonic matter 
extends over a large region of the computational domain.  
In contrast, the NSs shape is almost unchanged from its initial undisturbed state. 
At $t=10.8\ {\rm ms}$ (middle panel), the AS and NS have 
merged and the highest axionic and baryonic energy 
density centers around the origin of the numerical domain. 
A fraction of the material becomes unbound and leaves 
the central region.
At $t=17.6\ {\rm ms}$ (bottom panel) the remnant has a notable offset 
from the origin since the ejection of matter happens anisotropically 
such that the final object obtains a kick in the positive x-direction. 
The velocity of the remnant is of the order of $\sim 1000 {\rm km/s}$ and 
thus comparable with the observation of high velocity pulsars, e.g.~\cite{Arzoumanian:2001dv}. 

\section{The Multi-messenger picture of NSAS collisions}

\paragraph*{\textbf{Gravitational wave emission:--}}

\begin{figure}
\includegraphics[width=\columnwidth]{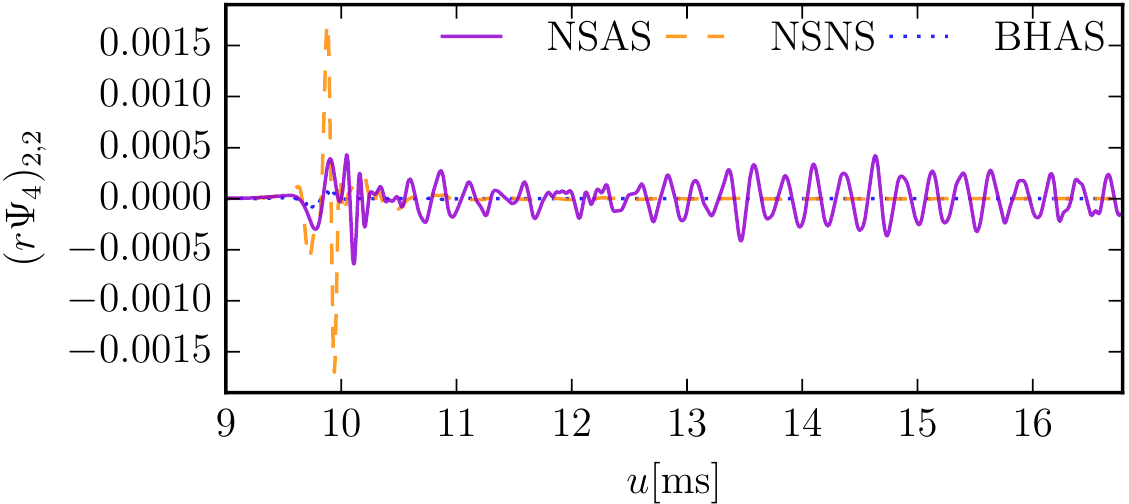}
\caption{Dominant (2,2)-mode of the GW signal for the NSAS simulation as well as a NSNS and BHAS employing the same individual masses shown for comparison.}
\label{fig:GW}
\end{figure}

In Fig.~\ref{fig:GW} we show for comparison the dominant 2,2-mode 
of the GW signal for the NSAS configuration as well as for a BHAS and NSNS
setup using the same masses for the binary constituents. 
We observe a clear ordering in the peak GW amplitude, 
which can be explained by the compactness of the individual objects. 
The smallest amplitude occurs for the BHAS merger, 
due to the large tidal distortion of the AS
in the BH's gravitational field, with  
axionic matter falling into the BH before the actual merger. 
This effect is reduced for the NSAS 
configuration due to the smaller compactness of the NS compared to the BH. 
For a BNS merger, we find that tidal deformations are 
significantly smaller again and consequently, the merger amplitude is the largest.
As is visible in the figure, the main differences 
in the GW emission with respect to a BNS system occur after 
the merger of the stars at frequencies for which current 
GW detectors are insensitive, see e.g.~\cite{Bezares:2017mzk}.
We find that GWs continue to be produced for the NSAS example case, 
such that the total energy emitted increases significantly over time. 
By the end of the simulation, the NSAS collision has released more energy in GWs 
than all the other cases, with a mass equivalent of the order 
$\mathcal{O}(10^{-3}M_\odot)$. 
Assuming a burst search with current GW detectors our example 
head-on NSAS merger would only be observable up to $\sim 100\ \rm kpc$ 
and for the Einstein Telescope (ET) up to 
$\sim 10\ \rm Mpc$~\citep{Abernathy:2011ET}.

Considering a potential quasi-circular NSAS merger, 
the emitted GW signal during the coalescence of a NSAS binary 
can be modelled in the same way as BNS systems and 
existing waveform models including tidal effects, 
e.g.~\cite{Hinderer:2016eia,Dietrich:2018uni,Nagar:2018zoe} 
are already in place for their construction
on quasi-circular orbits~\footnote{
We rely on~\cite{Cardoso:2017cfl,Sennett:2017etc}, who showed that 
it is possible to describe the deformation of exotic compact objects
within an external gravitational field with the help of tidal deformability 
parameters, similarly to the description of NSs.}. 
For such a quasi-circular merger, it would be possible to 
observe our example setup  up to 
$\sim 100\, \rm Mpc$ for advanced LIGO at design 
sensitivity~\citep{Martynov:2016fzi} 
and possibly $\sim1000\, \rm Mpc$ for ET~\citep{Punturo:2010zz}.

\paragraph*{\textbf{The Kilonova from baryonic mass ejection:--}}

\begin{figure}
\includegraphics[width=\columnwidth]{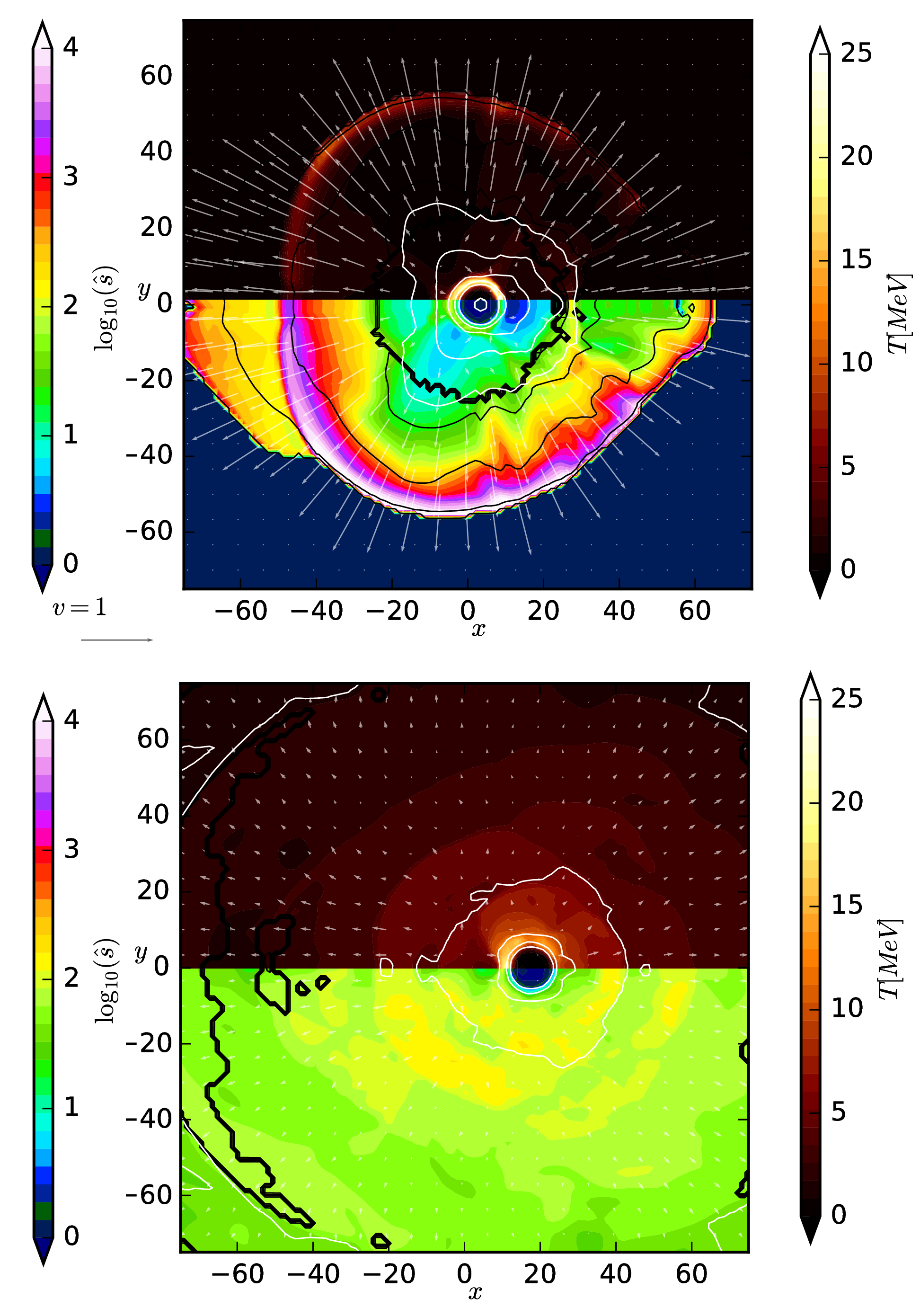}
\caption{Temperature (top part of the panels) and entropy (bottom part of the panels) for the times $t=10.8 \ {\rm ms}$ (top) and $t=17.6 \ {\rm ms}$ (bottom), cf.~Fig.~\ref{fig:dynamics}. White contour lines refer to the bound density, the black contours mark material which is unbound and ejected from the system.}
\label{fig:entropy}
\end{figure}

As shown in Fig.~\ref{fig:dynamics} 
a large fraction of bosonic and baryonic mass is ejected after the collision. 

In particular the baryonic ejecta may give rise to 
an optical transient similar to AT2017gfo. 
To test this hypothesis, we measure the amount of unbound baryonic
material and find that just $\sim 10 \ {\rm ms}$ after the collision 
$\sim 10^{-2}M_\odot$ of baryonic matter has been ejected. 
At the end of the simulation the ejection 
rate is still of the order of $\sim 2 \times 10^{-3}M_\odot/{\rm ms}$, i.e., ejecta will continue long after the simulated period. 
The main ejection mechanism for our configuration are shock waves formed during and after the merger of the two stars. 
In Fig.~\ref{fig:entropy} (top panel) we show the temperature and
entropy computed for the baryonic material. 
Both quantities can be estimated from our simulations, 
following~\cite{kyutoku2013black}. 
Thermal effects are included in our simulations using an 
additional thermal pressure  
$   p_{\rm th} = (\Gamma_{\rm th} -1 ) \rho \epsilon_{\rm th}$ 
where $\epsilon_{\rm th}$ denotes the thermal energy density 
and $\Gamma_{\rm th}$ the thermal adiabatic exponent, which we set 
here to $\Gamma_{\rm th}=1.75$; cf.~\citep{Bauswein:2010dn}. 
The entropy and temperature can then be estimated using
\begin{eqnarray}
    \hat{s} & = & \frac{p}{p_0}=\frac{p_0+p_{\rm th}}{p_0}, \\
    \epsilon_{\rm th} &=& \frac{3 k T }{2 m_u} + \frac{11 a_r T^4}{4 \rho},
    \label{eq:s}
    \label{eq:temp}    
\end{eqnarray}
where $k$ is the Boltzmann constant, 
$m_u$ the atomic mass unit, 
and $a_r$ the radiation-density constant. 

In Fig.~\ref{fig:entropy} (top panel) we can clearly 
see the shock wave leaving the central region. 
Furthermore, we find that around the central (cold) region of the 
NS, temperatures of the order of $\sim 20\ \rm MeV$ are reached. 
The temperature of the initial ejecta at the shock front is 
$\sim 10-15\ \rm MeV$. 
Because of this high temperature, 
the first ejecta component released immediately after the merger 
(top panel of Fig.~\ref{fig:entropy})
will have a high electron and consequently low lanthanide fraction. 
Ejecta released later have a lower temperature, 
but are still shock driven.

\begin{figure}
\includegraphics[width=\columnwidth]{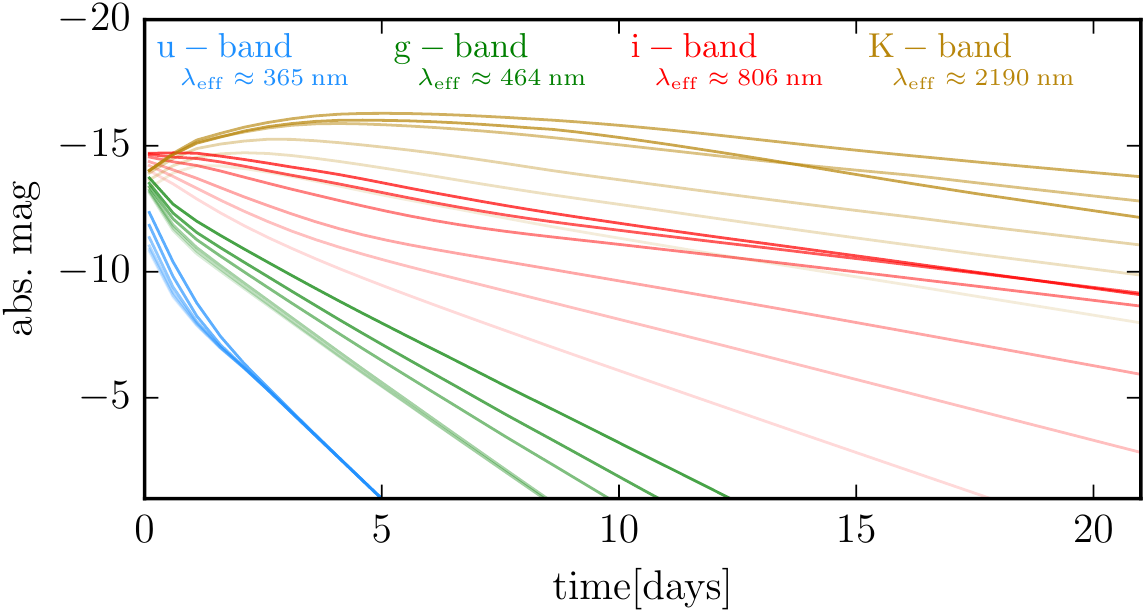}
\caption{Absolute magnitudes in the u-,g-,i-, K-band.
Increasing opacity indicates an increasing total ejecta mass. 
We assume for the initial shock ejecta released shortly
after merger an ejecta mass of 
$M_{\rm ej,1} = 5 \times 10^{-3}M_\odot$ and assume 
for the second ejecta component masses of 
$M_{\rm ej,2} = 5 \times 10^{-3}M_\odot, 1 \times 10^{-2}M_\odot, 
 2 \times 10^{-2}M_\odot,  5 \times 10^{-2}M_\odot,  
 1 \times 10^{-1}M_\odot,  2 \times 10^{-1}M_\odot, 
 4 \times 10^{-1}M_\odot$; see text for more details. }
\label{fig:lightcurves}
\end{figure}

To estimate the absolute magnitude of the transient based 
on the ejected baryonic material, we use the  
kilonova model of~\cite{Kasen:2017sxr,Coughlin:2018miv}. 
The model of~\cite{Kasen:2017sxr} employs a multi-dimensional 
Monte Carlo code to solve the multi-wavelength radiation 
transport equation for a relativistically expanding medium. 
We interpolate between the existing Monte Carlo simulations 
using the Gaussian Process Regression techniques discussed 
in~\cite{Coughlin:2018miv}. For our estimates we assume the first 
ejecta component to have a mass of 
$M_{\rm ej,1} = 5 \times 10^{-3}M_\odot$, 
an ejecta velocity of $v_{\rm ej,1} =0.45c$, 
and a lanthanide fraction of $X=10^{-3}$. 
For the remaining ejecta, we assume a velocity of 
$v_{\rm ej,2} =0.15 c$ and a lanthanide fraction of $10^{-2}$. 
Since we do not know the total ejecta due 
to the finite length of our simulation, we vary the ejecta mass as 
$M_{\rm ej,2} = 
 5 \times 10^{-3}M_\odot, 1 \times 10^{-2}M_\odot, 
 2 \times 10^{-2}M_\odot, 5 \times 10^{-2}M_\odot,  
 1 \times 10^{-1}M_\odot, 2 \times 10^{-1}M_\odot, 
 4 \times 10^{-1}M_\odot$ and present different 
 lightcurve estimates in Fig.~\ref{fig:lightcurves}. 
The largest absolute magnitudes are obtained in the near-infrared, 
cf.~i- and K-band, while in the optical and ultraviolet the signal is fainter. 
Assuming that we use the Zwicky Transient
Facility~\citep{Bel2014}, which has a limiting magnitude of 20.4 in 
the g-band in its nominal survey mode~\citep{Coughlin:2018lta}, then the
transient would only be observable for the first 
few days after merger up to a distance of $\sim 40\, \rm Mpc$. 
The Large Synoptic Survey Telescope ~\citep{Ivezic2014} at 
its anticipated limiting magnitude of 24.4 in the
g-band~\citep{Coughlin:2018lta} could observe the NSAS merger 
transient up to  $\sim 1.5\, \rm Gpc$. 

There are few wide-field surveys in the near-infrared, and so for comparison we instead highlight near-infrared imagers on large-aperture telescopes such as Gemini~\citep{2003PASP..115.1388H} with a limiting magnitude of around 22 in K-band in 30\,s exposures. Due to the larger magnitude in the near-infrared, the transient would be observable for $\sim 2$ weeks
up to a distance of $250\ {\rm Mpc}$. Similarly, with the Thirty Meter Telescope's infrared imaging spectrograph (IRIS)
\citep{WrBa2010}, with a limiting magnitude of around 27 in K-band in 900\,s exposures, the transient would be observable up to a distance of $\sim 2.5\ {\rm Gpc}$.

\paragraph*{\textbf{Radio flares from baryonic mass ejection:--}}

The interaction between the interstellar medium and the mildly and sub-relativistic 
baryonic outflows will generate synchrotron radiation, so called 
``radio flares'', (see e.g.~\cite{Nakar:2011cw}). 
Radio flares triggered by the ejected material are observable 
for months or even years after the merger.
For our NSAS configuration, 
the first ejecta 
component will lead to a peak in the radio band about 
$\sim 1\ {\rm year}$ after the merger with the radio 
fluence $\sim28\ {\rm Jy}/(D/{\rm Mpc})^2$. 
The more massive (but slower) ejecta component causes a peak in the radio band about $9$ years after the merger with a fluence of 
$\sim 6.6\ {\rm Jy}/(D/{\rm Mpc})^2$.
Comparing to the current capabilities of the Karl G. Jansky Very Large
Array (VLA) with limiting fluxes of 
$\sim10\ \mu {\rm  Jy}$, these objects are detectable out to 
$\sim 1\ {\rm Gpc}$.
With an approximately 100$\times$ higher sensitivity, the Square
Kilometer Array~\citep{Murphy:2009nt} will detect such signals 
out to $\sim 10\ {\rm Gpc}$.

\paragraph*{\textbf{Neutrino Emission because of shock heating:--}}

Due to the fact that our numerical simulations do not include a
full general relativistic radiation hydrodynamics scheme, 
we rely on simplified estimates to determine the neutrino luminosity. 
As shown in Fig.~\ref{fig:entropy} the merger remnant reaches 
temperatures much larger than the electron-positron pair production 
threshold ($\sim 0.511\ {\rm MeV}$). 
Consequently, positron captures 
($n + e^+ \rightarrow → p + \bar{\nu}_e$) lead to an increase 
of the electron fraction and the emission of 
electron-type anti-neutrinos. 

In general there are two different mechanisms which could heat 
up the NS remnant during the NSAS collision. 
First, the shocks created during the merger process. 
Assuming that the neutrino emission happens within the hot region 
shown in Fig.~\ref{fig:entropy} and that neutrinos 
with energies of $3 k T \sim 40\ {\rm MeV}$ are emitted, 
we obtain a neutrino luminosity of the order 
of $10^{51}\ {\rm erg/s}$.
A second mechanism would be the conversion of axionic matter 
within the neutron star into thermal energy, 
which is discussed further below. 

To estimate the amount of detectable neutrinos at detectors such as 
Super-Kamiokande~\cite{Fukuda:2002uc} we use the relation 
\begin{equation}
R_\nu = \frac{21.1}{\rm ms} 
\left[\frac{ 32\ {\rm kT}}{M_{\rm H_2O}}\right]    
\left[\frac{ L_{\rm nu}} {10^{53} {\rm erg/s}}\right]    
\left[ \frac{E_\nu } {15 \ {\rm MeV}} \right]
\left[ \frac{10\ {\rm kpc}} {D} \right]^2
\label{Eq:Rnu}
\end{equation}
as given in~\cite{Palenzuela:2015dqa}. 
We find that the detection of neutrinos 
created due to shock heating during the collision of an NSAS merger
will only be possible within our galaxy or in the near neighborhood 
of the Milky Way.

\paragraph*{\textbf{Neutrino emission because of axion-photon heating:--}}

While the multi-messenger channels discussed so far also exist 
for a typical BNS merger, the presence of axions allows for 
additional signatures. In particular, the presence of 
an external magnetic field (as commonly observed for NSs) may lead to axion-photon 
interconversion \citep{Raffelt:1987im}. While this process has been studied with respect to low mass ASs that may 
be abundant in the Milky Way halo
\citep{Barranco:2012ur,Iwazaki:2014wka,Bai:2017feq}, 
it has not been explored for a NSAS merger\footnote{We note that if the axion 
in question is the QCD axion, its coupling to gluons may also lead to additional energy dispersion into the NS.}. 

For the numerical relativity simulation, we assumed that the axion 
field was only coupled to baryonic matter via gravitational interaction, 
i.e., an interaction purely due to the mutual impact on the metric. Based on 
the results of these simulations, we now discuss the effects of adding the typical 
coupling between the axion and electromagnetism.\footnote{Note that as the axion photon 
coupling is topological, its effects do not change in a curved 
spacetime.} Accurate determination of these signals, discussed in this and the following section, 
would require full simulation of the relativistic axion-magnetohydrodynamics during the merger. 
We do not attempt to do this here. Instead, we seek to elucidate some of the possible effects, 
reviewing previously published approximations with the addition of some cautionary notes.

We first consider the effect of the AS in the bulk of the NS. The axion-photon interaction is described by the Lagrangian:
\begin{equation}
\mathcal{L} \supset - C \frac{\alpha}{f_a} \phi ~ {\bf E} \cdot {\bf B},
\label{mixing}
\end{equation}
with ${\bf E}$, ${\bf B}$, and $\alpha$ denoting the electric and 
magnetic fields and the fine structure constant, respectively. 
$C$ is a model dependent constant of $\mathcal{O} (1)$. 
This Lagrangian leads to a modification of Maxwell's equations and Gauss's Law becomes:
\begin{equation}
\label{Gauss}
\triangledown \cdot {\bf E} = \rho - C \frac{\alpha}{f_a} \triangledown \cdot (\phi {\bf B}).
\end{equation}

This leads to an axion-induced electric field in an electrically neutral medium ${\bf E}_{\phi} = -C \frac{\alpha}{f_a} \phi {\bf B}$, which creates a current ${\bf J}_{\phi} = \sigma {\bf E}_{\phi}$ in the NS (with the conductivity $\sigma$), as described in \cite{Barranco:2012ur}~.
This leads to a power dissipation into the NS: 
\begin{equation}
\label{bulkPower}
P =  \int \sigma E_{\phi}^2 d^3 x = \frac{C^2 \alpha^2 B^2 \sigma}{f_a^2} \int \phi({\bf x})^2 d^3 x.
\end{equation}
For our example NSAS merger the AS envelopes 
the NS and thus the integral is taken over the entire NS.
Note that for stable AS configurations, 
$\phi \propto f_a$, and therefore the power dissipated is independent 
of $f_a$. 
With $\phi \sim 10^{-5} M_{pl}$, which is the typical field 
amplitude within the NS region at the end of the simulation, we can 
restrict our analysis to the leading order term in Eq.~\eqref{mixing} 
and neglect the back reaction of the axion field onto the magnetic field since $\phi \ll f_a$. 
Taking a typical NS conductivity 
$\sigma \sim 10^{26} \, {\rm s}^{-1}$ \citep{Baiko:1995qg} we find:
\begin{dmath}
\label{bulkPowerNumbers}
P(t = 0) = 10^{57} \, {\rm GeV} \, {\rm s}^{-1} \times C^2 \left( \frac{B}{10^8 \, {\rm T}} \right)^2 \left( \frac{\sigma}{10^{26} \, {\rm s}^{-1}} \right) \left( \frac{R_{\rm NS}}{10 \, {\rm km}} \right)^3,
\end{dmath}
where $R_{\rm NS}$ is the NS radius, and $t = 0$ corresponds to the time of initial contact between the AS and the NS. In this case, the entire energy of the AS is released into the NS in $\sim 100\ \rm ms$.\\

However, we must also consider the dynamics of the growth of ${\bf E}_{\phi}$. Substituting Ohm's law into the axion-modified Maxwell equations, we obtain \cite{Long:2015cza}:
\begin{equation}
\label{eq:Egrowth}
\dot{\bf E} = \triangledown \times {\bf B} - C \frac{\alpha}{f_a} \dot{\phi} {\bf B} - C \frac{\alpha}{f_a} \triangledown \phi \times {\bf E} - \sigma {\bf E} - \sigma {\bf v} \times {\bf B},
\end{equation}
where ${\bf v}$ is the fluid velocity. We therefore find that the growth of ${\bf E}_{\phi}$ could be almost entirely suppressed by the NS's high conductivity, in the absence of compensating contributions from other terms in equation \eqref{eq:Egrowth}, as discussed in \cite{Ahonen:1995ky} for the case of axion dark matter in primordial magnetic fields. Furthermore, for a smaller $B$ or $C$ than assumed in equation \eqref{bulkPowerNumbers}, the energy dissipation would be slower. In this case, the AS may slowly heat the NS as they remain in contact after the merger. The energy released inside the bulk of the NS is unlikely to escape the NS directly as photons, but will increase the NS temperature, leading to increased neutrino emission.

In the case in which all the axions are converted, we obtain an increase of the NS's 
temperature of $\mathcal{O}(10^2\ {\rm MeV})$ 
which leads to a neutrino luminosity 
$10^{52} \ {\rm erg/s}$ of neutrinos with energy $\mathcal{O}(1\ {\rm GeV})$. 
Using Eq.~\eqref{Eq:Rnu} we find that a detection of neutrinos up to 
$\sim 1 \ {\rm Mpc}$ is possible if the emission happens over a time of $1$ second. Using the 
sensitivity of the future Hyper-Kamiogrande neutrino 
observatory~\citep{Nakamura:2003hk}, the neutrinos from axions 
conversion could be observable up to $\sim 20\ \rm Mpc$. 

\paragraph*{\textbf{Electromagnetic bursts from axion-photon interconversion:--}}
Unlike in the bulk, photons from axion to photon conversion in the
magnetosphere of the NS may be observed directly. Furthermore, the conductivity of the magnetosphere 
may be significantly lower than that of the bulk 
\citep{Li:2011zh}, and so the possible damping of ${\bf E}_{\phi}$ discussed above is avoided. We follow the approach 
of \cite{Iwazaki:2014wka} to estimate the size of this signal. 
As above, we have an axion generated electric field 
${\bf E}_{\phi} = \frac{-C \alpha \phi {\bf B}}{f_a}$. $\phi$ oscillates
with a frequency corresponding to the axion mass $m_a$, and so does 
${\bf E}_{\phi}$. 
The electrons in the magnetosphere therefore oscillate parallel to 
${\bf E}_{\phi}$, obeying the equation of motion $\dot{p} = e E$, 
with the electron's momentum $p$. 

The electrons may become highly relativistic during these 
oscillations~\cite{Melrose:2009jk,Reville:2010yy}. Their maximum possible $\gamma$ 
factor is $\gamma_{\rm max} = \frac{p_{\rm max}}{m_e} = \frac{e E_{\phi}}{m_a m_e} = 4.8 \times 10^6$. 
However, collisions between the electrons and other charged particles may reduce their maximum speed. 
If we assume the electron's speed returns to that determined by the local temperature after each collision, 
then relativistic speeds are obtained only at fewer than $10^{12}$ collisions per second. 
Furthermore, such collisions may result in the energy stored in the AS being dissipated as heat into the NS, 
rather than emitted as radiation, even in the magnetosphere.

The Larmor formula for radiated power 
$P = \frac{2 e^4}{3 m_e^2} E_{\phi}^2$ also applies for 
relativistic oscillations where acceleration is parallel to velocity.
The emitted radiation is strongly beamed along the direction of the
electron velocity, i.e. along the direction of ${\bf B}$. 
For dipole radiation for non-relativistic oscillating charged particles, 
the emitted radiation's frequency is the oscillation frequency. 
For axion species with masses below the mass considered here, 
this effect can be used to explain Fast Radio Bursts \cite{Iwazaki:2014wka}. In the relativistic case, radiation is emitted at all frequencies up to a cut-off frequency $\omega_c = \gamma_{\rm max}^2 m_a$.

Therefore, rather than giving Fast Radio Bursts, higher mass NSAS mergers may lead to emission of EM radiation at a continuous range of frequencies.

We perform an order of magnitude estimate of the radiated power taking $\phi \sim 10^{-5} M_{pl}$, and assuming collisions within the magnetosphere are not significant. Using the Larmor formula, the power radiated from a single electron is $P_1 = 10^{-3} \, {\rm GeV} \, {\rm s}^{-1} \times C^2 \left(\frac{B}{10^8 \, {\rm T}} \right)^2$. We take an electron density $n_e = 10^{24} \, {\rm cm}^{-3}$ \cite{Chamel:2008ca} and a magnetosphere depth $d = 1$ cm \cite{Ho:2009mm}.

\begin{dmath}
\label{magnetosphere}
P = 10^{33} \, {\rm GeV} \, {\rm s}^{-1} \times C^2 \left( \frac{B}{10^8 \, {\rm T}} \right)^2 \left(\frac{d}{1 \, {\rm cm}} \right) \left( \frac{n_e}{10^{24} \, {\rm cm}^{-3}} \right) \left( \frac{R_{\rm NS}}{10 \rm km} \right)^2.
\end{dmath}
In \cite{Iwazaki:2014wka}, it is argued that the axion field oscillates coherently within its de Broglie wavelength, and therefore the total power output is enhanced by an extra factor of $N_e$, the number of electrons in the de Broglie volume. This effect would lead to a factor of $10^{34}$ increase in the power output. However, the simulation is not sufficiently fine grained to test the coherence of the axion field, and furthermore the coherence of the oscillations could be destroyed by collisions within the magnetosphere. 
When the electrons obtain relativistic speeds, the majority of the emitted radiation is well above the plasma frequency of the magnetosphere, so we assume that a significant proportion will escape from the NS. The minimum duration of this EM burst will be $\mathcal{O}(100 \, {\rm ms})$, the duration of the merger. For lower power outputs, the EM radiation may also be released over a longer period as the AS remains bound to the NS. The NSAS merger could therefore generate a distinctive fast burst of continuum radiation.\\

The observability of axion to photon interconversion in the magnetosphere depends on 
several factors. In particular, the NS magnetic field and the properties of collision in the magnetosphere determine the frequency of the 
emission, and therefore which telescopes could observe it. For example, let us consider the case in which the emission peaks at optical energies. Neglecting collisions, this corresponds to $B \sim 3 \times 10^6$ T and $\omega_c = $ 2.1 eV. We optimistically assume a partial coherent enhancement of the signal, such that $P \propto N_e^{1.5}$, and assume that all the resulting power dissipation escapes the magnetosphere.

In this case we have a radiated power $P = 10^{47} \, {\rm GeV} \, {\rm s}^{-1}$. 
This power will be beamed along the direction of the magnetic field, which changes across 
the surface of the NS and with the NS's rotation. We therefore initially neglect the effect 
of beaming in this order of magnitude estimate. Assuming again the Zwicky Transient 
facility the burst of radiation lasting for about $100\ \rm ms$ might be seen up to 
a distance of $40\ \rm Mp$c.

On the other hand, we could consider the most optimistic case in which the NS's 
magnetic dipole is pointing directly at the Earth, with most of the magnetic field 
lines parallel to the line of sight throughout the NS's period. 
Taking $\gamma = 10^6$, the GRB is then tightly beamed within 
$\gamma^{-1}$ 
along the line of sight (LOS). 
In this case, the burst would be observable from a 
distance of $90\ \rm Gpc$. We note that for a dipole model of the 
NS's magnetic 
field the beaming is very tight, within a cone of angle $\sim 10^{-6}$, i.e., it is likely that the GRB emission would be 
entirely beamed away 
from the LOS and be invisible to us. 
However, in the case where the NS magnetic 
field has a more complex structure, or the beam rotates onto the 
LOS during the merger due to the NS's rotation, it may be observable. 
For the Large Synoptic Survey Telescope we obtain maximum observable
distances of $230\ \rm Mpc$ and $580\ \rm Gpc$ for the unbeamed 
and beamed scenario, respectively. 

\section{Summary and Outlook}
\label{sec:summary}

\begin{table}
  \centering    
  \caption{
  A wanted poster:  Multimessenger channels for a NSAS merger. 
  The columns refer to: the channel, 
  the observable distance with current 
  state-of-the-art techniques and the estimated observable distance 
  with techniques anticipated for the near future.} 
  \begin{tabular}{l|cc}        
    \hline
    Channel                        & $D_{\rm today}$        & $D_{\rm future}$ \\
    \hline
    GW$_{\rm head-on}$             & $\sim 0.1\rm {\rm Mpc}$ & $\sim 10\rm {\rm Mpc}$  \\
    GW$_{\rm inspiral}$            & $\sim 100\ {\rm Mpc}$ & $\sim 1000\ {\rm Mpc}$  \\    
    Kilonova$_{\rm u-band}$        & $\sim40\ {\rm Mpc}$    & $\sim1.5\ {\rm Gpc}$ \\
    Kilonova$_{\rm K-band}$        & $\sim250\ {\rm Mpc}$   & $\sim2.5\ {\rm Gpc}$ \\    
    Radio flare                    & $\sim1\ {\rm Gpc}$     & $\sim10\ {\rm Gpc}$ \\
    Neutrino$_{\rm shock}$         & $\sim 0.1\ {\rm Mpc} $ & $\sim 2\ \rm Mpc$ \\     
    Neutrino$_{\rm axion-heating}$ & $\sim1\ {\rm Mpc}$     & $\sim 20\ \rm Mpc$ \\
    EM burst$_{\rm axions-photon}^{\rm unbeamed}$ & $\sim 40\ {\rm Mpc}$   &  $\sim 90\ {\rm Gpc}$ \\
    EM burst$_{\rm axions-photon}^{\rm beamed}$ & $\sim 230\ {\rm Mpc}$   &  $\sim 580\ {\rm Gpc}$ \\
    \hline
  \end{tabular}
  \label{tab:observables}
\end{table}

We have quantified the possibility of observing 
our example NSAS merger with different multi-messenger channels,  
cf.\ Table~\ref{tab:observables}.

Interestingly, we found that a number of observables 
usually connected with BNS or NSBH mergers are also present for
NSAS mergers. This could lead to a misinterpretation of 
observable data, and so further studies identifying clear differences between BNSs and NSASs are required for 
a less ambiguous interpretation of future multi-messenger observations. Consider for example GW170817 -
the optical/infrared/ultra-violet signature of AT2017gfo is broadly consistent with the kilonova 
signature from NSAS systems, and the signal sGRB170817 could be
explained via formation of a BH and disk from a NSAS merger or
from axion-photon conversion. 
We have also shown that NSAS mergers can release up 
to $M_{AS} = 10^{56}$ GeV in energy in the optical band for
reasonable parameter values. Therefore, for
some choices of the NS conductivity and magnetic field strength, a 
NSAS merger could give rise to unusual transients such as
AT2018cow~\cite{Prentice:2018qxn}, in which an 
unexplained optical luminosity of 
$\sim 44 \, {\rm ergs} \, {\rm s}^{-1}$ was observed.

The confirmed detection of a NSAS merger would be a significant discovery, 
simultaneously confirming the axion as a dark matter component, 
and constraining its mass, decay constant, and couplings to 
standard model particles. 
It would thus provide a lead in our understanding of the
nature of dark matter within the Universe. 

\bibliographystyle{mnras}
\bibliography{paper20180814.bbl}

\section*{Acknowledgements}
\label{sec:acknowledge}

TD acknowledges support by the European Unions Horizon 2020 research and 
innovation program under grant agreement No 749145, BNSmergers.
Computations have been performed on the supercomputer SuperMUC at the LRZ
(Munich) under the project number pr48pu, and 
the compute cluster Minerva of the Max-Planck Institute for Gravitational Physics. 
This work has been partially supported by STFC consolidated grant ST/P000681/1. 
We thank David J E Marsh and MC David Marsh for useful discussions.
Furthermore, TD and KC want to thank S.Khan for helpful 
comments on our simulations of ASs.

\bsp    
\label{lastpage}
\end{document}